\documentclass[twocolumn,prl,showpacs]{revtex4}
\usepackage[pdftex]{graphicx}
\usepackage{amsmath}
\usepackage{graphicx}
\usepackage{times}

\begin{document}

\title{Ionic correlations at the nanoscale: inversion of selectivity in a bio-nanochannel.}
\author{Marcel Aguilella-Arzo}
\affiliation{Biophysics Group, Department of Physics, Universitat Jaume I, 12080 Castell\'o, Spain.}
\author{Carles Calero and Jordi Faraudo}\email{jfaraudo@icmab.es}
\affiliation{Institut de Ci\`encia de Materials de Barcelona (ICMAB-CSIC), Campus de la UAB, E-08193 Bellaterra, Spain}
\begin{abstract}
Here we show, combining a simulation and theoretical study, that electrostatic correlations typical of multivalent ions can reverse the selectivity of a biological nanochannel. Our results provide a physical mechanism for a new, experimentally observed phenomenon, namely the inversion of the selectivity of a bacterial porin (the
\emph{E. Coli} OmpF) in presence of divalent and trivalent cations. Also, the differences and similarities between the driving force for this phenomenon and other similar nano and micro-escale electrokinetic effects (e.g. inversion of streaming current in silica nanochannels) are explored.
\end{abstract}

\date{\today }
\maketitle
\begin{figure}[htp]
  \begin{center}
     \includegraphics*[width=4cm]{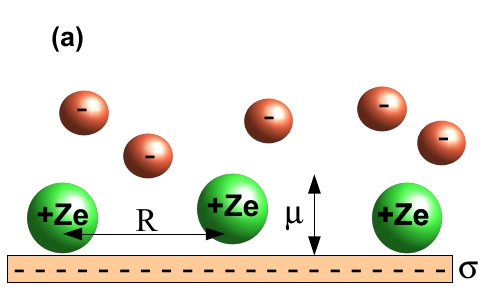}
     \includegraphics*[width=4cm]{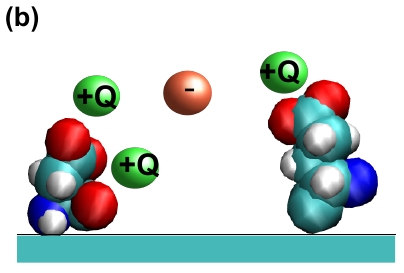}
      \caption{\label{correlations}Ionic correlations near charged interfaces. (a) High electrostatic coupling \cite{Grosberg2002,Levin2002,Netz2005}. Counterions of charge $+Ze$ accumulate near a strongly charged interface within a thin layer of typical thickness $\mu=k_BT\epsilon/2\pi(-\sigma)Ze$ of the order of the ionic size. At this layer, the cations have strong positional correlations, leaving correlation holes of radius $\pi R^2\sim Ze/(-\sigma)$ ($R\gg\mu$ so the coupling parameter is large $\Gamma\equiv R/2\mu\gg1$ ) (b) Correlations between ions and a surface fuctionalyzed with charged groups \cite{Faraudo2007,Calero2010}. Here, correlations result from the electrostatic interactions between multivalent positive counterions (green) and negatively charged oxygen atoms (red) located in interfacial groups. In this case, the direct interaction between charged interfacial atoms and counterions is essential.}

   \end{center}
\end{figure}

In recent years, the development of statistical-mechanical theories of ionic correlations near strongly charged interfaces \cite{Grosberg2002,Levin2002,Netz2005,Kjellander2009} allowed a unified physical understanding of new phenomena arising in electrokinetic processes as diverse as ionic transport in nanochannels \cite{Heyden2006,He2009,Labbez2009}, DNA condensation \cite{Gelbart2000,Besteman2007} or particle electrophoresis \cite{Jonsson2005,MartinMolina2008}. 
 Theories typically focus on the high electrostatic coupling regime, appearing in the case of multivalent ions and highly charged interfaces \cite{Grosberg2002,Levin2002,Netz2005,Kjellander2009}. In this regime, ion-ion correlations are largely independent of the chemical nature of the interface so the physics of the problem can be understood using simple models for the interface and primitive models for the electrolyte \cite{Grosberg2002}. A different theoretical proposal shows the possibility of strong electrostatic correlations between multivalent ions and charged chemical groups located at the interface \cite{Vaknin2005,Faraudo2007,Calero2010}. These correlations are an interfacial analogue of the well-known Bjerrum correlations appearing between multivalent ions in bulk electrolyte\cite{Vaknin2005}. They are relevant for interfaces containing (or covered by) well-separated charged interfacial groups which interact \emph{individually} with ions in solution (see Figure \ref{correlations}). It is important to note that these ion-interfacial group correlations do not require a highly charged interface and they depend strongly on the nature of the structure and charge distribution of the chemical groups present at the interface.


\begin{figure*}[htp!]
   
  \begin{center}
      \includegraphics*[width=13cm]{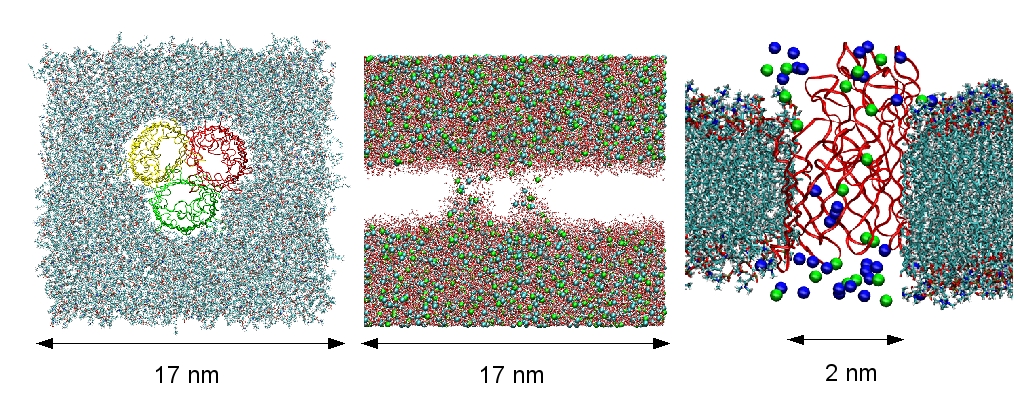}
      \caption{\label{Fig:Snapshot}\textbf{Snapshots of the simulated system from different views}. (a) top view of the trimeric OmpF protein inserted in a lipid bilayer. The structure of the protein has been constructed from the X-ray crystallographic data \cite{Cowan} and using the protonation states described by Varma\cite{Varma} for the titrable residues (the resulting overall charge is $-11e$ per monomer)  (b) Front view of a snapshot showing water and ions (MgCl$_2$ in this case). The spacing between the two water slabs corresponds to the lipid membrane and the protein (not shown here for clarity). (c) Zoom showing one of the channel pores and ions located inside the pore and in nearby solution (for clarity of the representation, all water molecules, many lipids and the other two monomers are not shown). The picture has been produced using VMD\cite{VMD}. }
   \end{center}
\end{figure*}

In this letter, we will show how correlations of the type described in Fig. \ref{correlations}b are behind a new experimental phenomena, namely the inversion of selectivity observed in a biological nanochannel\cite{Alcaraz2009,Alcaraz2010} in presence of multivalent cations. This biological nanochannel is the so-called OmpF ionic channel found in the outer membrane of \emph{E. coli}. This is a relatively wide ionic channel in the sense that it allows the simultaneous permeation of both cations and anions (in \emph{hydrated} form) at high rates. This channel is made of three identical monomerical nanopores (see Figure \ref{Fig:Snapshot}), each one leaving an hourglass-shaped aqueous pore with a diameter between 1-4 nm surrounded by many charged titratable residues with an overall negative charge \cite{Aguilella2007} at $p$H=7. In presence of monovalent electrolyte, it has a slight cationic selectivity (i.e. the flux of cations is larger than the flux of anions across the pore) which is well understood from basic electrostatic concepts \cite{Aguilella2004,Aguilella2007}. Unexpected experimental results\cite{Alcaraz2009,Alcaraz2010} show that the cationic selectivity of the channel found for 1:1 electrolytes is reversed for sufficiently high concentrations of 2:1 or 3:1 electrolytes. The different divalent cations (Mg$^{2+}$, Ca$^{2+}$, Ba$^{2+}$, Ni$^{2+}$) give almost identical results, whereas trivalent cations (La$^{3+}$) induce a substantially larger effect. Another interesting piece of evidence is provided by X-ray\cite{XrayMg} data obtained from crystalls of OmpF protein channels in 1 M of MgCl$_2$. In this structure, hydrated Mg$^{2+}$ ions were found in contact with certain negatively charged residues, suggesting a strong cation-interfacial group interaction. All this evidence (strong dependence on valency but small dependence on the particular cation, strong role of particular chemical groups) suggests the existence of important ion-interfacial group correlations of electrostatic origin in this system.

In order to identify the basic physical mechanism underlying the observed selectivity inversion, we have performed all-atom molecular dynamics (MD) simulations with NAMD2 \cite{NAMD2} of the OmpF channel in different electrolytes (see Fig.\ref{Fig:Snapshot}). In order to study the transport properties of the channel, we have applied an external electric field  of $14.22$ mV/nm perpendicular to the lipid bilayer, which creates an electrostatic potential drop of $\approx 200 $ mV across the membrane+channel system (see Supporting Online material). This value of the potential drop was selected in order to be high enough to ensure ion permeation across the channel during the simulation and also because it is experimentally accessible. It has to be stressed that full-atomistic MD simulations of transport in protein channels are extremely challenging, becoming possible only due to recent improvements in algorithms and computer power \cite{Aksimentiev2005,Ulrich2009}. In this case, we have conducted the first simulations of ionic transport in a protein channel in divalent electrolyte, which extraordinarily increases the need for longer simulation times and larger simulation systems, as discussed in detail in Ref \cite{molsim}. All simulation details, algorithms, processing of the data and simulation movies can be found in the online supplementary information \cite{EPAPS}.

\begin{table}
\caption{Ionic fluxes (number of ions observed to cross the OmpF channel during production runs of duration $t_{\text{run}}$) and occupancy numbers (defined as number of ions of each species inside the protein channel averaged over the simulation runs, statistical errors in the mean are estimated from $2\sigma$). The results correspond to simulations in three different conditions: 1 M KCl, 1 M MgCl$_2$ and a mixture of 1 M MgCl$_2$ and 1 M KCl.}
\label{Table:Fions}
\begin{tabular}{|l|c|ccc|ccc|}
\hline
           & $t_{\text{run}}$ & \multicolumn{3}{|c|}{Ionic Flux} & \multicolumn{3}{c|}{Channel Occupancy} \\
        & (ns) & Cl$^-$ & K$^+$ & Mg$^{2+}$ & Cl$^-$ & K$^+$ & Mg$^{2+}$\\ \hline
     KCl & 24.9 & 38 & 47 & - & $21.0 \pm 0.2$ & $35.5 \pm 0.2$ & - \\
     MgCl$_2$ & 36.8 & 32 & - & 1 & $47.1 \pm 0.9$ & - & $33.9 \pm 0.3$ \\
     Mixture & 31.6 & 35 & 14 & 7 & $24.4 \pm 0.2$ & $7.5 \pm 0.2$ & $7.2 \pm 0.2$ \\
\hline  
    \end{tabular} 
\end{table}

\begin{figure*}[htp!] 
      \includegraphics[width=7.5cm]{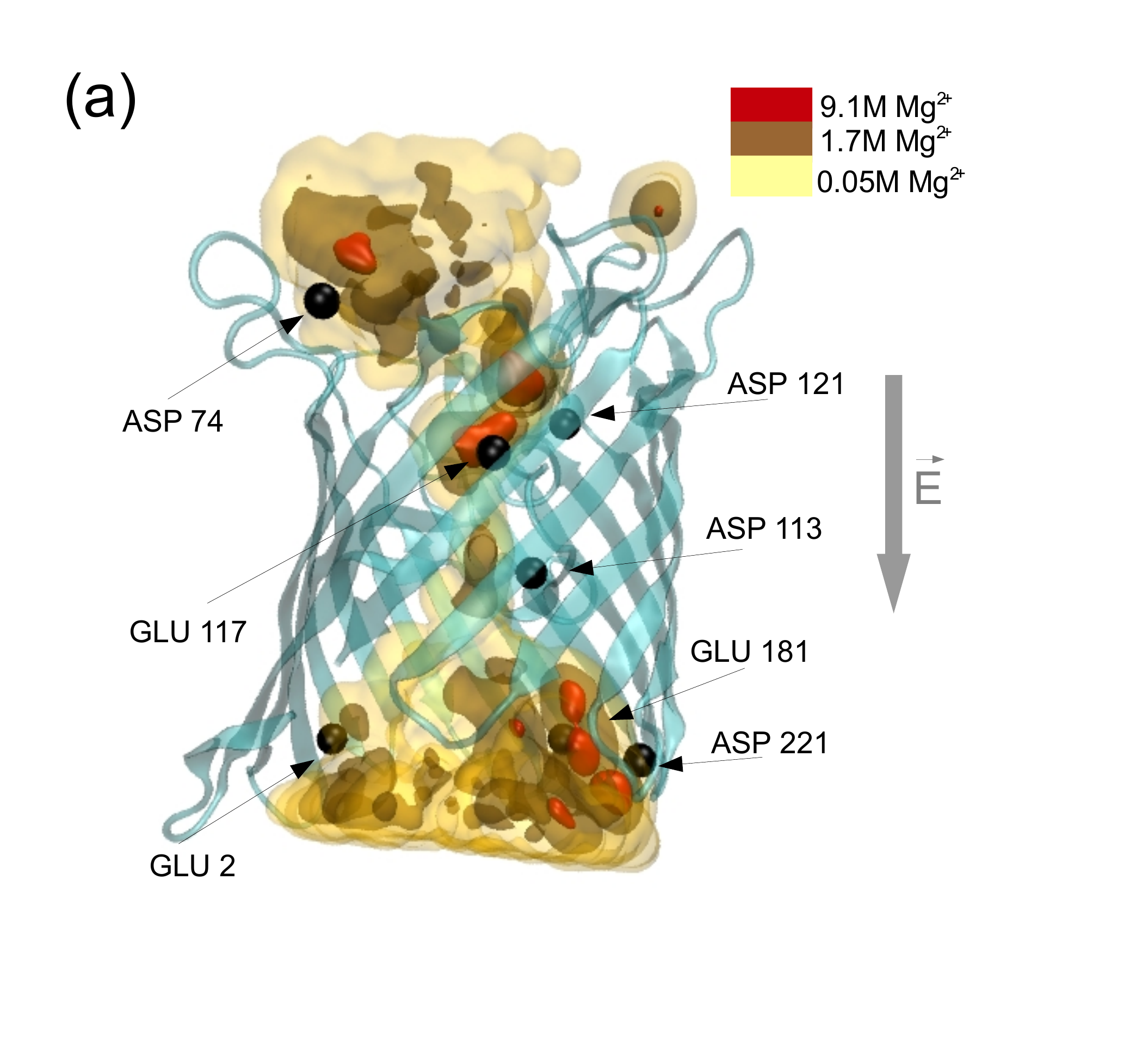}
      \includegraphics[width=10cm]{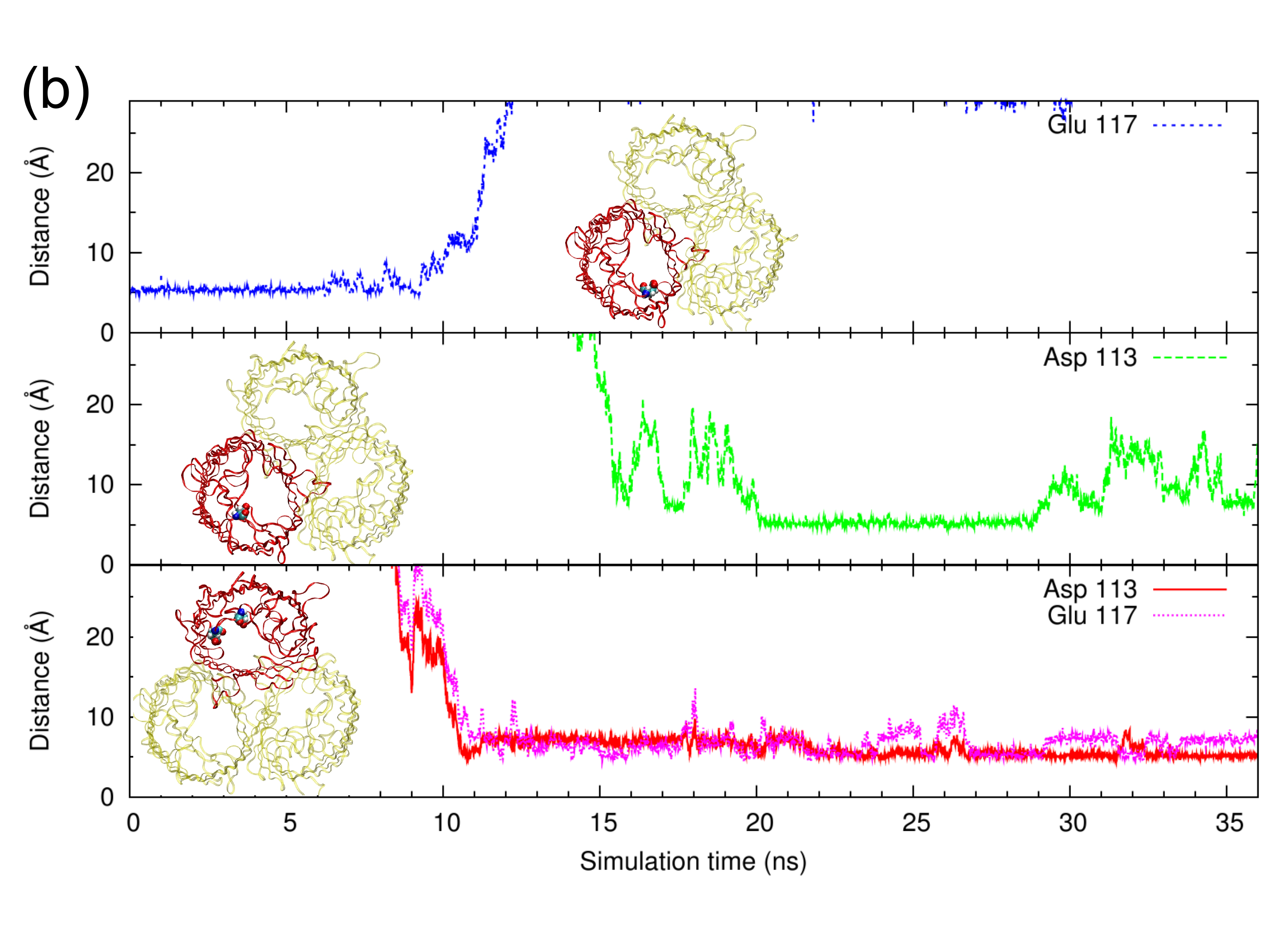}
      \caption{ \label{Fig:Isodensity} Interaction between Mg$^{2+}$ cations and the protein channel in simulations with 1 M MgCl$_2$. (a) Concentration of Mg$^{2+}$ inside one of the monomers of the protein channel. We show three different isosurfaces averaged over a 10 ns trajectory fragment of the production run. Strong inhomogeneities (with concentrations 1 order of magnitude larger than bulk concentration) are found near negative groups located at the protein surface, schematically indicated as black balls (picture produced with VMD \cite{VMD}) (b) Examples of long residence times of Mg$^{2+}$ ions near negatively charged residues evidenced by the time evolution of the distance between three selected Mg$^{2+}$ ions and certain protein residues indicated in the onsets. Top panel: ion close to the residue Glu117 of monomer P2 at the start of the production run which remains there roughly 10 ns. Medium panel: ion initially at bulk solution enters inside the P2 monomer of the channel and remains in contact with the Asp113 residue during 9 ns. Bottom panel: ion initially at bulk solution enters inside the P3 monomer of the channel and remains in contact with both the Asp113 and Glu 117 residues (no detachment was observed).}      
\end{figure*}

%

First of all, let us note that our results for the transport of KCl across the channel (see Table \ref{Table:Fions}) are in agreement with previous work\cite{Roux,Aguilella2007,Varma,Ulrich2009}. The flux of K$^+$ is larger than that of Cl$^-$ and the average number of K$^+$ inside the channel is larger than the average number of Cl$^-$ (Table \ref{Table:Fions}), as expected for a negatively charged channel.

The situation is completely different in the case of MgCl$_2$ (see Table \ref{Table:Fions}). The channel has inverted its selectivity: a channel conducting a slightly cationic current (in the case of KCl) has turned into an almost purely anion conducting channel in the case of MgCl$_2$. Let us remark that, as expected for a negatively charged channel, there is an excess of cations in its interior as compared with bulk electrolyte (we observe a ratio of 0.72 between the average numbers of Mg$^{2+}$ and Cl$^-$ ions compared with 0.5 in bulk electrolyte). In spite of the excess of cationic charge inside the channel, its contribution to the current is extremely low and the vast majority of the observed current is due to  Cl$^-$ . We also recall that in simulations\cite{molsim} of 1 M bulk solutions of MgCl$_2$, 43$\%$ of the current is due to Mg$^{2+}$, so the effect observed here has to be attributed to the protein channel.  The lack of flow of the many Mg$^{2+}$ cations populating the channel is due to the strong, attractive interactions with channel walls. The strong correlations between the positions of Mg$^{2+}$ cations and negatively charged acidic residues of the channel can be clearly appreciated in Figure \ref{Fig:Isodensity}. Mg$^{2+}$ dwell near the strong concentration peaks observed in Fig.\ref{Fig:Isodensity}a and they are hardly found in the rest of the channel. In fact, residence times of Mg$^{2+}$ ions near negatively charged groups could be very large (of the order of 10 ns), as shown in the examples in Figure \ref{Fig:Isodensity}b. In table \ref{Table:binding} we show the average number of Mg$^{2+}$ in close proximity of negatively charged acidic groups located at the protein walls. Note that some of these acidic groups have an average number of Mg$^{2+}$ larger than 0.5, indicating that the cationic charge overcompensates the -$e$ charge of the acid. Therefore, we observe local \emph{charge inversion} \cite{Faraudo2007}.

The strong correlations found between Mg$^{2+}$ ions and acidic residues located at the channel's wall can be considered the first clear evidence for the transversal correlations predicted by recent statistical-mechanical theories of ionic correlations\cite{Vaknin2005,Faraudo2007,Calero2010}. In these theories, the electrostatic interaction between an ion of charge $q_C$ and an interfacial atom of charge $q_I$ generates an excess concentration of the ion near the interfacial charge characterized by a pairing (or binding) constant $K_I$ given by: 
\begin{equation}
\label{Bjerrum}
 K_I \simeq 2\pi \int_{d_0}^{d_c} r^2 e^{q_Iq_C l_B/r}  dr.
\end{equation}
In Eq.(\ref{Bjerrum}), $l_B=e^2/4\pi\epsilon k_BT$ is the so-called Bjerrum length (0.714 nm for water at 25C), $d_0$ is the distance of closest approach between ions and interfacial charged atoms and $d_c=q_C q_I l_B/2$ is the typical correlation length between the ions and the interfacial charges. In our simulations, the interfacial atoms are oxygen atoms from acidic residues (which have typical partial charges $q_C=-0.7e$). For Mg$^{2+}$, we obtain $d_c\approx0.5$ nm. In our simulations, Mg$^{2+}$ is found to retain their hydration water and a distance of closest approach about $d_0\approx 0.4$ nm is observed, so we obtain $K_I\approx0.7$ M$^{-1}$. In the case of K$^+$, we obtain $d_c\approx 0.25$ nm which is smaller than the sum of the crystallographic radius of oxygen and K$^+$ (0.28 nm), hence correlations are negligible and $K_I$ vanishes. It is also interesting to estimate $K_I$ for trivalent cations. In the case of La$^{3+}$, we have $d_c\approx 0.75$ nm. Since this ion has a size close to Mg$^{2+}$ and also tends to remain hydrated, we can also take $d_c\approx 0.4$ nm obtaining a very high affinity of $K_I\approx 6.4$M$^{-1}$. Experimental results\cite{Alcaraz2010} show that the inversion of selectivity for OmpF is larger in presence of LaCl$_3$ than in the MgCl$_2$ case, as expected from our calculations.

\begin{table*}
 \caption{Average number of Mg$^{2+}$ found nearly in contact to certain acidic residues (see Figure 3), evaluated by averaging the number of Mg$^{2+}$ cations at distances between $d_0\approx 0.4$ nm and $d_c\approx0.5$ nm  of oxygen atoms of the acidic residues (see the main text) during all production runs. We show the results for simulations with 1 M of MgCl$_2$ and a mixture of 1 M of MgCl$_2$ and 1 M KCl (statistical errors in the mean are estimated from $2\sigma$.)}
\label{Table:binding}
\begin{tabular}{|c||c|c|c|c|c|c|c|}
\hline
         & Glu2 & Glu181 & Asp221 & Asp74 & Asp121 & Asp113 & Glu117 \\
\hline
MgCl$_2$ & $0.51 \pm 0.05$& $0.71 \pm 0.05$ & $1.04 \pm 0.07$ & $0.28 \pm 0.03$ & $0.66 \pm 0.05$ & $0.40\pm 0.05$ & $0.33 \pm 0.03$ \\
Mixture  & $0.44 \pm 0.06$& $0.48\pm 0.06$ & $0.65 \pm 0.06$ & $0.31 \pm 0.06$ & $0.74\pm 0.08$& $0.50 \pm 0.06$ & $0.44\pm 0.05$ \\
\hline
\end{tabular}
\end{table*}
Further insight on the physical basis of this inversion phenomenon can be obtained by analyzing the effect of adding a high amount of screening monovalent salt to a situation with inverted selectivity. The two different kinds of ionic correlations illustrated in Figure \ref{correlations} behave in opposite ways under the addition of high amounts of background monovalent electrolyte. Experiments in silica nanochannels found selectivity inversion\cite{Heyden2006,He2009} in presence of MgCl$_2$ but the original selectivity of the channel is recovered\cite{Heyden2006} (no inversion) in mixtures of multivalent electrolyte and KCl. This finding is in agreement with theoretical predictions\cite{Heyden2006} based on ion-ion correlations between multivalent ions near strongly charged interfaces (see Figure \ref{correlations}a). On the other hand, theories predict that the short range electrostatic correlations between multivalent ions and interfacial groups (see Figure \ref{correlations}b) are essentially not affected by the addition of 1:1 electrolyte. In this case, one expects that the original selectivity is not recovered after the addition of 1:1 salt; rather one may expect to observe a \emph{weakened} inverted selectivity. Actually, this is what we observe in our simulations of the ion channel bathed by a mixture of 1M KCl and  1 M MgCl$_2$ (see Table \ref{Table:Fions}). The strong transversal correlations between Mg$^{2+}$ cations and acidic groups are also observed, with a similar intensity to that observed in the previous case with only MgCl$_2$ (see table \ref{Table:binding} and the isodensity figure available as Online Supporting Material).The observed anionic current is larger than the cationic current: the initial cationic selectivity of the channel is not recovered. The cationic current obtained with 1 M of MgCl$_2$ and 1 M of KCl is \emph{significantly smaller} than that obtained in presence of only 1 M of KCl. Also, both the occupancy number and flux of Cl$^-$ observed in this case are close to the values observed in the case with only KCl, in spite of having 3 times more anions in the electrolyte solution (see table \ref{Table:Fions}). Hence, we can say that selectivity effects are weakened by the addition of 1:1 salt (thus demonstrating their electrostatic origin) but the inversion of selectivity has not been suppressed by adding 1:1 salt. This different response to added 1:1 electrolyte between our system and silica nanochannels emphasizes the different nature of the ionic correlations found in each case. 

Overall, our calculations provide a strong evidence showing the role played by electrostatic correlations in the selectivity of ionic transport in a wide biological nanochannel. In the view of recent nanotechnological applications of biological nanochannels and their engineered versions (modified by directed mutagenesis) \cite{Nature_nanotech} it is tempting to suggest that our results may have also implications in the design of responsive channels for biotechnological applications. From the point of view of biophysics, it is also interesting to note that recent electrophoretic results in phospholipid vesicles \cite{Faraudo2010} also show a complex behavior in presence of multivalent ions which is robust against addition of 1:1 electrolyte. All these evidences suggest that biologically relevant molecules (lipids or proteins) have mechanisms to preserve a strong interaction with multivalent ions even in presence of physiologically amounts of screening salt.

\section*{Acknowledgments}
This work is supported by the Spanish Government (grants FIS2009-13370-C02-02, FIS2007-60205 and CONSOLIDER-NANOSELECT-CSD2007-00041), Generalitat de Catalunya (2009SGR164) and Fundaci\'o Caixa Castell\'o-Bancaixa (P1-1A2009-13). C.C. is supported by the JAE doc program of the Spanish National Research Council (CSIC). The Supercomputing resources employed in this work were provided by the CESGA Supercomputing Center, Spain. We acknoweledge useful discussions with V.M. Aguilella, A. Travesset, R.S. Eisenberg, P. Carloni and R. Kjellander.
\\


\begin{thebibliography}{99}
\bibitem{Grosberg2002}A.Y. Grosberg, T.T. Nguyen and B.I. Shklovskii, Rev. Mod. Phys. \textbf{74}, 329 (2002). 
\bibitem{Levin2002}Y. Levin, Rep. Prog. Phys. \textbf{65}, 1577 (2002).
\bibitem{Netz2005}H. Boroudjerdi et al., Phys. Rep. \textbf{416}, 129 (2005).
\bibitem{Kjellander2009}R. Kjellander, J. Phys:Condens. Matter  \textbf{21} 424101 (2009).
\bibitem{Heyden2006}F.H.J. van der Heyden et al., Phys. Rev. Lett. \textbf{96}, 224502 (2006).
\bibitem{He2009}Y. He, Y et al., J. Am. Chem. Soc. \textbf{131}, 5194 (2009).
\bibitem{Labbez2009}C. Labbez, B. J\"onsson, M. Skarba and M. Borkovec, Langmuir \textbf{25}, 7209 (2009).
\bibitem{Gelbart2000}W.M. Gelbart, R.F. Bruinsma, P.A. Pincus and V.A. Parsegian, Phys. Today {\bf 53}, 38 (2000).
\bibitem{Besteman2007}K. Besteman, K. van Eijk and S. Lemay, Nature Physics \textbf{3}, 641 (2007).
\bibitem{Jonsson2005}B. J\"onsson et al., Langmuir {\bf 21}, 9211 (2005).
\bibitem{MartinMolina2008}A. Mart\'in-Molina, J.A. Maroto-Centeno, R. Hidalgo-\'Alvarez and M. Quesada-P\'erez, Colloids and Surfaces A \textbf{319}, 103 (2008).
\bibitem{Vaknin2005}A. Travesset and D. Vaknin, Europhys. Lett. \textbf{74}, 181 (2006).
\bibitem{Faraudo2007}J. Faraudo and A. Travesset, J. Phys. Chem. C \textbf{111}, 987 (2007). J. Faraudo and A. Travesset, Biophys. J. {\bf 92}, 2806 (2007).
\bibitem{Calero2010}C. Calero and J. Faraudo, J. Chem. Phys. \textbf{132}, 024704 (2010).
\bibitem{Alcaraz2009}A. Alcaraz et al. Biophys. J. \textbf{96}, 56 (2009).
\bibitem{Alcaraz2010}E. Garc\'ia-Jim\'enez, A. Alcaraz and V.M. Aguilella, Phys. Rev. E \textbf{81} 021912 (2010).
\bibitem{Aguilella2007}M. Aguilella-Arzo et al. Bioelectrochem. \textbf{70}, 320 (2007).
\bibitem{Aguilella2004}A. Alcaraz et al. Biophys. J. \textbf{87}, 943 (2004).
\bibitem{XrayMg}E. Yamashita et al. EMBO J \textbf{27}, 2171 (2008).
\bibitem{NAMD2}J.E. Phillips et al., J. Comp. Chem \textbf{26}, 1781 (2005)
\bibitem{molsim}C. Calero, M. Aguilella-Arzo and J. Faraudo, J. (in preparation).
\bibitem{EPAPS}See supplementary material at http://link.aps.org/supplemental/.
\bibitem{Aksimentiev2005}A. Aksimentiev and K. Schulten, Biophys. J. {\bf 88}, 3745 (2005).
\bibitem{Ulrich2009}S. Pezeshki et al. Biophys. J., \textbf{97}, 1898 (2009).
\bibitem{Cowan}S. W. Cowan et al., Structure {\bf 3}, 1041 (1995).  
\bibitem{Varma}S. Varma, S. Chiu and E. Jakobsson, Biophys. J. {\bf 60}, 112 (2006).
\bibitem{Roux}W. Im and B. Roux, J. Th. Biol. \textbf{322}, 851 (2002).
\bibitem{VMD}W. Humphrey, A. Dalke and K. Schulten, Molec. Graphics {\bf 14.1}, 33 (1996)
\bibitem{Nature_nanotech}V.M. Aguilella and A. Alcaraz, Nature Nanotechnology \textbf{4}, 403 (2009).
\bibitem{Faraudo2010}A. Mart\'in-Molina, C. Rodr\'iguez-Beas and J. Faraudo, Phys Rev. Lett. \textbf{104} 168103 (2010).
\end{thebibliography}
\end{document}